\documentclass[options]{acmart}  

\makeatletter
\def\@ACM@checkaffil{
    \if@ACM@instpresent\else
    \ClassWarningNoLine{\@classname}{No institution present for an affiliation}%
    \fi
    \if@ACM@citypresent\else
    \ClassWarningNoLine{\@classname}{No city present for an affiliation}%
    \fi
    \if@ACM@countrypresent\else
        \ClassWarningNoLine{\@classname}{No country present for an affiliation}%
    \fi
}
\makeatother

\AtBeginDocument{%
  \providecommand\BibTeX{{%
    \normalfont B\kern-0.5em{\scshape i\kern-0.25em b}\kern-0.8em\TeX}}}

\usepackage{algorithm}
\usepackage{algorithmic}

\usepackage{siunitx}
\usepackage{multirow}
\usepackage[T1]{fontenc}    
\usepackage{hyperref}       
\usepackage{url}            
\usepackage{booktabs}       
\usepackage{amsfonts}       
\usepackage{nicefrac}       
\usepackage{microtype}      
\usepackage{lipsum}
\usepackage{subfigure}
\usepackage{graphicx}
\usepackage{amsmath,bm}
\usepackage{xcolor}         

\usepackage{algorithm}
\usepackage{algorithmic}

\setcopyright{acmcopyright}
\copyrightyear{2023}
\acmYear{2023}
\acmDOI{XXXXXXX.XXXXXXX}

\acmPrice{15.00}
\acmISBN{978-1-4503-XXXX-X/18/06}

\def \xx  {{\bf x}}
\def \vv  {{\bf v}}
\def \bb  {{\bf b}}
\def \WW  {{\bf W}}
\def \hh  {{\bf h}}
\begin{document}

\title{Personalized Search Via Neural Contextual Semantic Relevance Ranking}  

\author{Deguang Kong, Daniel Zhou, Zhiheng Huang and Steph Sigalas}
\email{ doogkong@gmail.com, {danzhou, zhiheng, stephsig}@amazon.com}
\affiliation{Amazon}
\authornote{The work was done while authors were/are working at Amazon. Correspondence to: Deguang Kong<doogkong@gmail.com>.}
\renewcommand{\shortauthors}{D. Kong, et al.}

\begin{abstract}
Existing neural relevance models do not give enough consideration for  query and item context information
which diversifies the search results to adapt for personal preference.  To bridge this gap, this paper presents a neural learning framework to personalize document ranking results by leveraging the signals to capture how the document fits into users' context. In particular, it models the relationships between document content and user query context using both lexical representations and semantic embeddings such that the user's intent can be better understood by data enrichment  of personalized query context information. Extensive experiments performed on the search dataset, demonstrates the effectiveness of the proposed method. 
 
\end{abstract}






\maketitle
\section{Introduction}

Search personalization refers to the presentation of personalized search results based on the individual user accessing the ranking result. Search engines adopt contextual information~\cite{DBLP:conf/sigir/JiangKCC14} relevant to user intent and query context, to improve the ranking results and reduce the ambiguity. Since both the query context and document context reformulation are important indicators of context information in ranking query and document pairs, we argue that modeling this information would be beneficial to personalized search tasks.  For example, the ranking of items should be increased if they are more relevant to the context of a search query. Suppose an Engineer in the USA enters a search query of ``benefits'' into the search interface, then a document with the relevant context of ``engineer" and ``USA" will be ranked higher. 

In this work, we propose a neural learning framework to increase document ranking relevance based on document context. We model the document context information by matching it to the user context information in queries, where the commonality between user query context and document content is explicitly modeled to capture their interactions over in search sessions. To summarize, we list the key contribution of this work  as follows. 

\begin{figure}[t]
  \centering
  \includegraphics[width=0.48\textwidth]{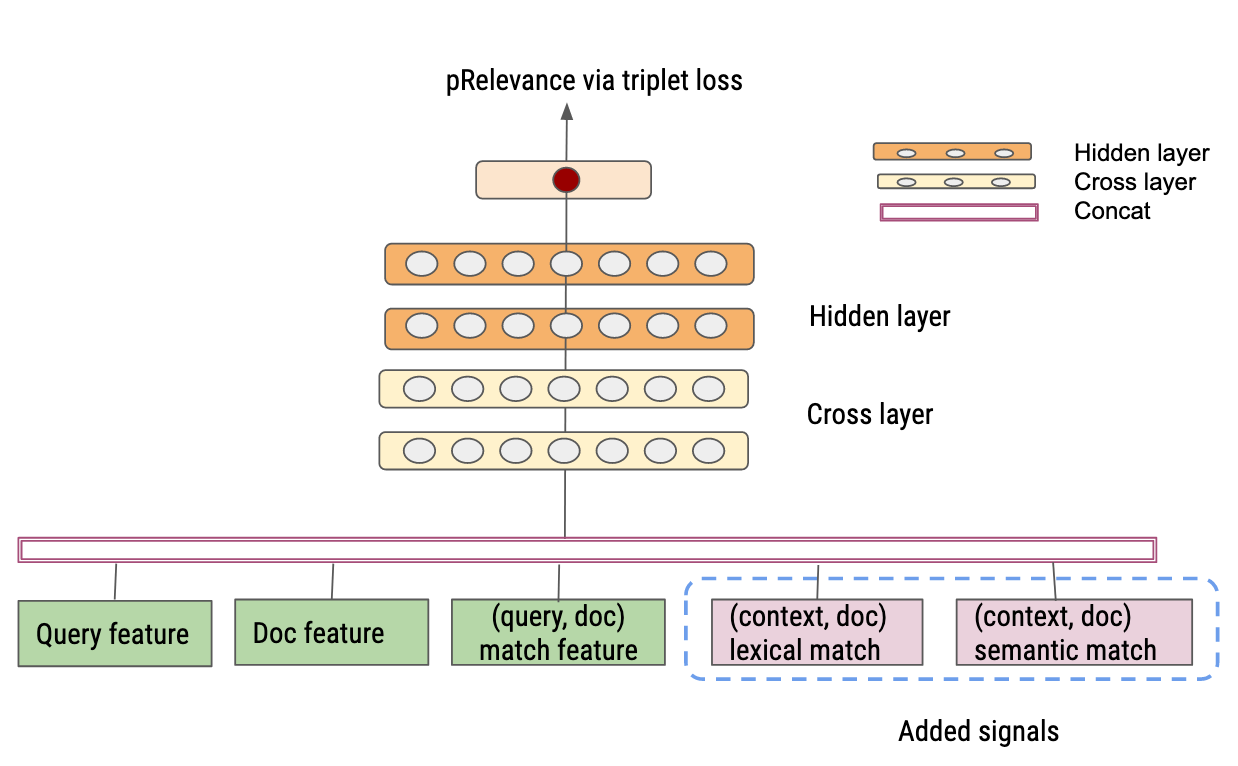}
  \caption{\small Personalized Search via neural contextual semantic relevance ranking in the LTR framwork with deep cross network for modeling pRelevance score between (query, doc) pairs by considering (context, doc) relevance using triplet loss. \\}
  \label{fig:dcn}
\end{figure}

\begin{itemize}
\item We present a personalized LTR framework based on contextual enrichment via data augmentation that allows to incorporate both document context and user query context information.

\item To the best of our knowledge, we provide the \emph{first} benchmark search dataset that leverages the document's contextual information for improving the search quality, based on  human annotations to facilitate the work along this direction. 


\item The document context and user query context information are interacted properly in a holistic way to improve rank relevance, with demonstrated performance gains over baseline methods. 
\end{itemize}

\section{Neural 
Ranking Framework}

In the paper next, we define the query as $q$ submitted by a user with a specific search intent. Every query $q$ 
is associated with a set of related documents $D = \{D_1, \cdots, D_m\}$ that are ranked by its
relevance to the query, and $Y = \{y_1, \cdots, y_m\}$ is the set of relevance labels for each document in $D$. In a typical search engine, $y_i$ is usually modeled by a categorical variable, i.e., \{Prefect, Good, Fair, Bad\}. A query $q_i$ generally consists of a short sequence of words as $q_i = \{q^1_i, q^2_i, \cdots, q^n_i\}$ and document $D_j$ consists of title and body sequence and $D_j = \{D^t_j, D^b_j\}$. The query context is denoted as a set of attributes $C = \{C_1, C_2, \cdots, C_K\}$, e.g., geo, job family and etc.  

{\bf Problem Definition} The context relevance ranking task studied in this paper refers to the rank of the searching result based on their relevance w.r.t the given queries by considering user intent and query context. We not only have to consider the relevance between the document and the query, but also wish that the higher-ranking documents are correlated with the context of the query such that the search engine provides personalized ranking results based on user query context.   The key \emph{challenge} is to maintain the semantic consistency between the surfaced document and the query context. In this paper we focus on explicit context that describes users’ segmentation information (e.g., geo and job family) clearly at user-cohort level (instead of introducing  vagueness or ambiguity).   Prior IR approaches (~\cite{liu2022category},~\cite{fan2022modeling},~\cite{wang2022efficient}) do not give enough considerations for explicit context at user cohort level, although many researches have been performed for penalization of search results based on user interaction behaviors~\cite{DBLP:conf/wsdm/KocayusufogluWS22}, such as click-steam and conversion channels.  In contrast, this paper presents a method to 
adapt the ranking results based on how the document fits both users' intent and underlying context information.

\vspace{-3mm}
\subsection{High-level Idea}
Ranking the retrieved document for an input query and its context is the problem we wish to solve. More formally, let $Pr(D|q,C)$ be the relevance score between the document and the input query and its associated context and this can be formulated as 
\begin{eqnarray}
 Pr(D|q, C) \propto  Pr(D|C) Pr(D|q) Pr(q, C),  
\label{Eq:pr_dqc}
\end{eqnarray}
where $Pr(D|q)$ models the traditional ranking relevance ~\cite{DBLP:conf/cikm/HuangHGDAH13} between document and query, $Pr(D|C)$ models how the document fits the context, and $Pr(q, C)$ gives the prior information about how the query is associated with the particular context (which is fixed given the specific query and context). The final ranking score should combine the document-query relevance $Pr(D|q)$, document-context ranking score  $Pr(D|C)$ based on prior distributions of query and context pairs $Pr(q, C)$. 

{\bf System Workflow} 
Given a search query from the search session, e.g., ``benefit'', the system will first capture the context of a search query. The query interpretation automatically interprets the operators and filters in the user’s query. In particular, the contexts would be a set of named attributes for a specific search query. For example, an Engineer in Seattle entered a search query  ``benefit", the  context attribute of the query would be {``engineer", ``Seattle"}. It is evident that in current search ranking results, this context information has not been necessarily met in the learning-to-rank (LTR) results. A  straightforward idea is to capture document context and see how the document is relevant to the user’ query context.  For example, we may check ``an employee benefit document" and see whether it is relevant to the context of {``engineer", ``Seattle"}. However, the document context relevance score is missing  in many documents. Therefore, a contextual-semantic matching component is needed to capture the document context relevance score. After obtaining this score, we integrate this score into a standard LTR framework for improving the search quality.  

\vspace{-3mm}
\subsection{Neural Contextual Semantic Ranking}
\label{section:neural_context_semantic_ranking}
The core idea of neural contextual semantic relevance ranking is to predict the relevance score between each query context and document corpus, which we define as \emph{document-context relevance score}. More formally, for each context attribute $k$, it would need to model the relevance $\mathcal{S}^k (C_j, D_i)$ between a document $D_i$ and context value $C_j$ for each attribute category $k$, i.e, 
\begin{eqnarray}
\label{Eq:s_(ci,dj)}
Pr(D_i|C_j)  \propto  \mathcal{S}^k (C_j, D_i).
\end{eqnarray}
The signals can be extracted via lexical representations or semantic representations. In practice, we combine them together to take advantage of each individual strength at both lexical granularity and semantic granularity levels.  

\begin{figure}[t]
  \centering
  \includegraphics[width=0.50\textwidth]{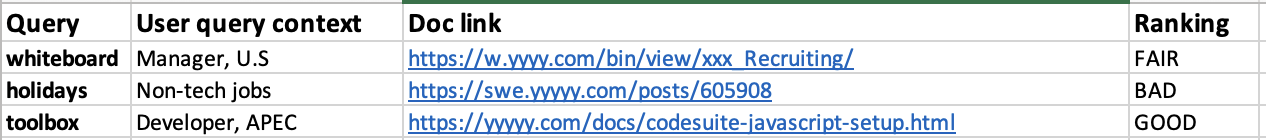}
  \caption{\small An example of annotated personalized search dataset given (query, doc) pairs with extra user query context information (the doc websites are anonymized). \\}
  \label{fig:dataset}
\end{figure}

{\bf Lexical representations}
One straight-forward way of computing Eq(\ref{Eq:s_(ci,dj)}) is using lexical representation of both context and documents to capture the matching information at token-level. Basically,  it
heuristically combines token overlap information, from which they compute a
matching score for context and document pairs. Given its popularity in existing systems, we would adopt BM25~\cite{DBLP:journals/ftir/RobertsonZ09} as a candidate. Given a context $c$ and document $d$, it will generate a score based on overlapping token statistics between context-document pairs, i.e, 
\begin{eqnarray}
\label{Eq:s_bm25}
 \mathcal{S}_{lex} (c, d) =  \sum_{t \in {c  \cap d}}  \text{r}_t  \frac{tf_{c,d}}{ tf_{c,d} + k_1[(1-b) + b \frac{|d|}{\ell}]},
\end{eqnarray}
where $t$ is a term, $tf_{t,d}$ is $t$'s frequency in document $d$,  $r_{t}$ is the $t$'s Robertson-Sparck
Jones weight~\cite{robertson1976relevance}, $\ell$ is the average document length, and $k_1$ and $b$ are parameters.

{\bf Contextual Semantic embedding}
The semantic embedding model can encode both the context ($c$) and document $d$ information into the dense embedding vectors (i.e., $v_{c}\in \Re^{d}$, $v_d \in \Re^{d}$) before computing their similarity in the embedding space.  Instead of using CNN, LSTM~\cite{10.1162/neco.1997.9.8.1735} architectures, we leverage the pre-trained SentenceBERT~\cite{reimers-gurevych-2019-sentence} model to generate the embeddings by average pooling representations from the encoder's last layer, i.e., 
\begin{eqnarray}
\label{Eq:sentence_bert}
{\bf v}_c = avgPooling(Bert_{\theta} (context)), \; 
{\bf v}_d = avgPooling(Bert_{\theta} (document)) \nonumber 
\end{eqnarray}
The context-document matching score  $\mathcal{S}_{sem} (c, d)$ is defined as the dot-product of embedding vectors of ${\bf v}_c$ and ${\bf v}_c$ as it allows accelerations using vector quantization~\cite{pmlr-v119-guo20h} for efficient feature computations, i.e., 
\begin{eqnarray}
\label{Eq:s_bert}
 \mathcal{S}_{sem} (c, d) =  \frac{ {\bf v}^{\intercal}_c  {\bf v}_d}{ \|{\bf v}_c\|  \|{\bf v}_d \| }.
\end{eqnarray}

\subsection{End to End optimization}

Fig.~\ref{fig:dcn} gives an overview of the LTR framework using deep cross network, which consists of feature extraction and modeling part. In the feature extraction stage, we stack the existing features extracted from query $q$ and documents $D=\{d_i\}$ side, along with the document-context $c$ matching features (illustrated in Section ~\ref{section:neural_context_semantic_ranking}) into the dense feature representations, i.e., 
\begin{eqnarray}
\label{Eq:x_feature}
\xx(q,d,c) = [ \vv_{query}, \vv_{doc}, \vv_{qMd}, \mathcal{S}_{lex} (c, d), \mathcal{S}_{sem} (c, d)], 
\end{eqnarray}
where $\vv_{query}$, $\vv_{doc}$, $\vv_{qMd}$ denotes query features, document features, and document-query matching features typically used in search ranking system, $\mathcal{S}_{lex} (c, d)$ and $\mathcal{S}_{sem} (c, d)$ are the contextual features extracted from Eq.(\ref{Eq:s_bm25}) and Eq.(\ref{Eq:s_bert}), respectively.

Since deep cross network~\cite{10.1145/3442381.3450078} can learn feature interactions automatically to capture feature interactions, we adopt DCN model and feed $\xx(q,d,c)$ to it to generate the feature embeddings by emphasizing the feature interactions among document-context matching score and other features, which actually maps input $\xx(q,d,c)$ to embeedings in the last hidden of $\ell$ layer ($ F(q, d, c) \overset{\Delta}{=} \hh_{\ell}$) (please refer to Appendix~\ref{section:DCN} for details).

{\bf E2E optimization} For E2E optimization, given the set of the query, documents, and human-labeled task-specific data $\{q, D=\{d_i\}, Y = \{Y_i \in [0, 1, 2, 3]\}\}$, we adopt a \emph{triplet loss} an an objective to minimize: 
\begin{eqnarray}
\label{Eq:ps_loss}
\mathcal{L^{\text{hinge}}}(q,D,Y) = \sum_{q} \sum_{i, j}  {\rm I}(y_i > y_j)  \max \big[ 0, \zeta - (F(q,d_i) - F(q, d_j)) \big]
\nonumber 
\end{eqnarray}
where $\rm I(y_i > y_j)$ is an indicator function that maps elements of the subset to one if the rank of document $y_i$ is larger than $y_j$ given query $q$ and all other elements to zero,  $\zeta$ is the parameter tuned in hinge loss (typically set to 1.0) which indicates the margin enforced between positive and negative pairs, and $F(q,d_i)$ is the semantic score learned using DCN from Eq.(\ref{Eq:DCN}).  
In optimization, the model was trained end-to-end and we used mini-batch SGD with Adam~\cite{DBLP:journals/corr/KingmaB14} for optimization.

\begin{table}[!t]
\small
\caption{Dataset description from two domains (D1 and D2)}
\label{tbl:dataset}
\begin{center}
\begin{tabular}{ p{3cm}|p{3cm} p{2cm}p{1cm}  p{1cm} }
   \toprule
{\bf Dataset} & 
{\bf domain}  &
{\bf \# query} & 
{\bf \# docs} &
{\bf contextual signals }\\ 
\midrule
D1-A &  1   &  266 & 89k & w/ \\
\midrule 
D1-B & 1  &  288 & 399k & w/o \\
\midrule 
D2-A & 2   &  5193 & 3213k  & w/  \\
\midrule
D2-B & 2  &   5193 & 3213k & w/o \\
\bottomrule
\end{tabular}
\end{center}
\end{table}

\begin{table}[!t]
\small
 \caption{Model performance on D1-A testing dataset}
  \centering
  \begin{tabular}{l|l|cccc}
    \toprule
Training data  & context features   & ndcg@10  & MAP & p@10   &  recall@10 \\
    \midrule
D1-A  &    w/ & 0.5882 &  0.3945  & 0.4414  & 0.442\\
D1-A &  w/o   & 0.0550  & 0.0480 &  0.0602 &  0.2101  \\ 
mixed training &  w/  & 0.5791 &  0.3873  & 0.4375  &0.4390 \\ 
mixed training & w/o  & 0.0483 &  0.0432 &  0.0602  & 0.2056 \\
    \bottomrule
  \end{tabular}
  \label{tab:D1A}
\end{table}

\begin{table}[!t]
\small
 \caption{Model performance on D1-B testing dataset}
  \centering
  \begin{tabular}{l|l|cccc}
    \toprule
Training data  & context features   & ndcg@10  & MAP & p@10   &  recall@10 \\
    \midrule
D1-B & w/  & 0.5003 & 0.4587 &  0.3950 &  0.7737 \\
D1-B & w/o  & 0.5003 &  0.4587 &  0.3950 &  0.7737 \\
mixed training  & w/ & 0.5071 & 0.4610 &  0.3963  & 0.7728 \\
mixed training & w/o & 0.5071 & 0.4610 &  0.3963 &  0.7728 \\
    \bottomrule
  \end{tabular}
  \label{tab:D1B}
\end{table}

\section{Experiment Results}

We conducted experiments on the collected search dataset using an intelligent enterprise search service that allows users search across different content repositories given built-in connectors.

\subsection{Dataset benchmarking} 

Since there is no ready personalized data set that incorporates user query context and doc context, we build benchmark datasets for personalized search. In particular, we collected datasets from two industry search applications, where domain 1 was from a big tech company\footnote{Due to privacy concerns, we are restrained from revealing more details of the datasets.} and domain 2 was from an insurance company, as summarized in Table~\ref{tbl:dataset}. Each domain consists of two datasets, one with contextual signals and the other w/o contextual signals. 

For the dataset w/o contextual signals, we have features (refer to Eq.\ref{Eq:x_feature}) generated from (query, doc) pairs and obtain relevance labels such as \{perfect, good, fair, bad\}.  For the dataset w/ contextual signals, we generate (context,doc) features in addition to (query, doc) features.  The relevance labels are annotated by annotates as \{perfect, good, fair, bad\} to indicate how the document is relevant to the queries by considering users' contextual signals as well (Fig.\ref{fig:dataset} gives an example). The average length of the queries used in the experiment is around 5.6, and the maximum allowable number of retrieved documents is set to 500.

\vspace{-3mm}
\subsection{Experiment settings and results} 


We train the model using D1-A, D1-B dataset respectively. For each dataset, we divided the data into
training and test sets, with the percentage of 80\%, 20\%  respectively. Since the D1-B dataset does not contain any contextual signals, we perform \emph{mixed training} by combining D1-A, D1-B dataset together where contextual signals are set to be zeros for D1-B dataset. 
We test the model performance on D1-A and D1-B datasets, whose performance are presented in Table~\ref{tab:D1A} and Table~\ref{tab:D1B}, respectively.  
\begin{table}[!t]
\small
 \caption{Generalization Capability: model performance on D2-A testing dataset}
  \centering
  \begin{tabular}{ll|cccc}
    \toprule
Training data  & context features   & ndcg@10  & MAP & p@10   &  recall@10 \\
    \midrule
D2-A & w/  & 0.4414 & 0.4042  & 0.0332& 0.8067 \\
D2-B & w/o & 0.2600 & 0.2233& 0.0241& 0.6238 \\
D1-A + D1-B & w/ & 0.4351 & 0.3972 &0.0330 &  0.8044   \\  
    \bottomrule
  \end{tabular}
  \label{tab:generalization_A}
\end{table}

\begin{table}[!t]
\small
 \caption{Generalization Capability: model performance on D2-B testing dataset}
  \centering
  \begin{tabular}{ll|cccc}
    \toprule
Training data  & context features   & ndcg@10  & MAP & p@10   &  recall@10 \\
    \midrule
D2-A & w/o  & 0.3243 & 0.2765 & 0.0306& 0.7884 \\
D2-B & w/o & 0.3243 & 0.2765 &  0.0306 &  0.7884 \\
D1-A + D1-B & w/o & 0.3146  & 0.2631  &0.0298&  0.7693  \\ 
    \bottomrule
  \end{tabular}
  \label{tab:generalization_B}
\end{table}

{\bf (1)} After adding the contextual signals, the ranking performance has been significantly improved on D1-A dataset (shown in Table~\ref{tab:D1A}) both for in-domain data training using only D1-A data and mixed training with both D1-A dataset and D1-B dataset. This demonstrates the effectiveness of adding contextual signals, which  also implies the strong correlations between the relevance score and contextual signals. 

{\bf (2)} The relevance ranking performance is neutral when compared mixed training (using both D1-A and D1-B dataset) (shown in Table~\ref{tab:D1A} and Table~\ref{tab:D1B}) against single-dataset training on both D1-A and D1-B datasets. This indicates we are able to serve the model from mixed training for traffic w/ and w/o contextual signals, but without introducing any performance loss.  

{\bf Generalization capability} To show how the model can be transferred to out-of-domain data, we collect another dataset D2-A, D2-B from domain 2, which has no overlap of queries and docs with domain 1.  Similarly, D2-A dataset provides contextual signals, whereas D2-B is absent of such signals.  We use the model trained from domain 1 (with mixed training) to test model performance on domain 2. Table~\ref{tab:generalization_A} and ~\ref{tab:generalization_B} present the performance comparisons. We observe that the model can generalize well from domain 1 to domain 2 (with slight performance loss). 


\subsection{Ablation study} 
{\bf Impact of lexical features vs. semantic features} In the model training, we incorporate both lexical feature of Eq.(\ref{Eq:s_bm25}) and semantic feature of Eq.(\ref{Eq:s_bert}) since semantic matching features can be complementary to the lexical features which perform exact token matching but can not handle vocabulary mismatch very well.  Table~\ref{tab:feature_combination} 
shows the experiment results using only lexical features and semantic features for training the model in mixed training on D1 dataset. We observe the performance gains by combining both lexical granularity and semantic granularity features on other datasets as well.


\begin{table}[!t]
\caption{NDCG@10 at D1-A datasets}
\label{tbl:dataset}
\begin{center}
\begin{tabular}{ p{3cm}|p{3cm} p{2cm} }
   \toprule
{\bf Dataset} & 
{\bf features} &
{\bf NDCG@10}   \\ 
\midrule
mixed training &  lexical only &  0.5478  \\
\midrule 
mixed training & semantic only  &  0.5691   \\
\midrule 
mixed training & combination   & 0.5882    \\
\bottomrule
\end{tabular}
  \label{tab:feature_combination}
\end{center}
\end{table}

{\bf Impact of loss functions and semantic embeddings} We investigated the role of loss functions and pre-trained sentence-BERT embeddings.  We changed the pairwise hinge loss  
 to pairwise pairwise logistic loss of  Eq.\ref{Eq:ps_log_loss}), but only found subtle performance changes (i.e., ndcg@10 changed from 0.4351 to 0.4346 on D2-A using mixed training). We found slight performance differences using different versions of sentence-BERT embeddings  (i.e., $\thicksim$0.005 absolute changes in ndcg@10). However, we found significant performance drop (i.e., $\thicksim$0.15 absolute changes in ndcg@10)  if we do not adopt any pre-trained sentence-BERT embeddings. 

\vspace{-3mm}
\section{Related Work}


{\bf Document Ranking and Ad-hoc Retrieval} Traditional lexical based methods perform exact matching of query and document words with different normalization and weighting mechanism includes BM25~\cite{DBLP:journals/ftir/RobertsonZ09}, query likelihood~\cite{DBLP:conf/sigir/PonteC98}, etc. Deep neural network based document ranking methods firstly embed the queries and documents into dense representation space, and the ranking is calculated based on queries, document embeddings and other relevant features such as DRMM~\cite{DBLP:conf/cikm/GuoFAC16}, DSSM~\cite{DBLP:conf/cikm/HuangHGDAH13}, etc.  In addition, the interactions between query embedding and document embedding are considered in ~\cite{DBLP:conf/www/Mitra0C17}. 
Recently, pre-trained language (PLM) models~\cite{DBLP:conf/naacl/DevlinCLT19} have shown state-of-the-art performances ~\cite{DBLP:journals/corr/abs-1901-04085}~\cite{DBLP:journals/corr/abs-1910-14424} for ranking the document. Reconciling PLM-based ranking’s efficiency and effectiveness is a critical problem in real-world deployment since the computation cost generally scales quadratically to the input text length. For example, ColBERT~\cite{DBLP:conf/sigir/KhattabZ20} introduces the late interaction layer to model the fine-grained query-document similarity between the query and the document using BERT, Pyramid-ERNIE~\cite{DBLP:conf/kdd/LiuLCSWCY21} architecture exploits the noisy and biased
post-click behavioral data for relevance-oriented pre-training using BERT. However, none of these works give sufficient considerations for query context and document context, which is thoroughly studied in this work based on PLM models.

{\bf Contextual Search} Contextual search~\cite{DBLP:conf/cikm/KraftMC05} is a type of web-based search that optimizes the searching results based on the context provided by  the user.  For example,  in enterprise level search engine(e.g., ~\cite{DBLP:conf/iui/LuPLW11}), the query context can be derived from certain job-related user properties (e.g. job title, function, department, etc.) or are already managed in IT systems like directory services. In addition, the physical condition that user used to enter the query,  time related factors (e.g, season/trend),  user previous search queries/experience,  building off of previous knowledge  that allows queries to be automatically augmented for similar contexts (in a session or across-session), user profile/interest would be obtained based on particular user queries. It is recognized that the search history~\cite{hailpern2011youpivot} and contextual relations~\cite{lykke2021role} play important roles in enterprise search. In customer search engine, many strategies have been applied to personalized search result based on mining the rich query logs, including historical clicks~\cite{DBLP:conf/www/DouSW07}, user interest~\cite{DBLP:conf/www/QiuC06}, query-session information~\cite{DBLP:conf/cikm/ShenTZ05}, friend network~\cite{zhou2021group}, etc. Compared against these existing works, this paper provides a new angle of incorporating query context information (in the form of user attribute) by modeling the document-context relevance, which provides additional signals for optimizing the ranking results.

\vspace{-3mm}
\section{Conclusion}
In this work, we propose a personalized search ranking framework with data enrichment of contextual signals, and show that incorporation of the contextual signals can benefit document ranking tasks. This paper builds  the benchmark datasets (with human annotations) to show the effectiveness of personalized search with incorporated personalized contextual signals.  As our future work, we would like to leverage the personalized contextual signals to benefit Q\&A tasks.

\bibliographystyle{ACM-Reference-Format}
\bibliography{sample-base}

\appendix

\section{Appendix}
\label{section:ablation_study}
\subsection{Evaluation metrics}


{\bf Evaluation metrics} For the document ranking task, we need to rank the most relevant document in the descending orders, we use several standard ranking metrics, including mean average precision (MAP), Normalized Discounted Cumulative Gain (NDCG), precision@k and recall@k (the precision and recall values obtained for top $k$ documents existing after each relevant document is retrieved) to rank the position of documents.  All training used a mini-batch size of 32 that would be fit in GPU. Learning rate was set to 0.001. The code is implemented in Python/Mxnet and the training was performed on GPU machines,  where the algorithm converges to the minimum loss on the validation set.



 \subsection{DCN model details}
 \label{section:DCN}
 Deep Cross Nets (DCN) actually maps input $\xx(q,d,c)$ to embeedings in the last hidden of $\ell$ layer ($ F(q, d, c) \overset{\Delta}{=} \hh_{\ell}$)
, i.e.,  
\begin{eqnarray}
\label{Eq:DCN}
\text{Cross layer:} \;\;\;\;  
\xx_0 :&=& \xx(q,d,c) \nonumber \\
\xx_{\ell+1} &=& \xx_0 \odot (\WW_{\ell} \xx_{\ell} + {b_\ell} ) + \xx_{\ell}   \\
 \ell &=& 1, 2, \cdots, L    \nonumber  \\
 \text{Hidden layer:} \;\;\;\; 
\hh_\ell &=& \xx_{\ell}  \nonumber \\
\hh_{\ell+1} &=& f(\WW_{\ell} \hh_{\ell} + b_{\ell}) \;\;  (\ell = L+1,\cdots, L+k)  \nonumber \\
F(q, d, c) :&=&  \hh_{\ell+1} 
\end{eqnarray}
where $\xx_0 \in \Re^p$ is the input feature vectors, $\xx_{\ell}, \xx_{\ell+1} \in \Re^{p}$ represents the input and output of the $\ell+1$-th cross layer in DCN,  $\WW \in  \Re^{p \times p}$ and $\bb \in \Re^p$ denote the learned weight matrix and bias vectors respectively, $\hh_{\ell}$ and $\hh_{\ell+1}$
denote the input and output of the $h$-th hidden layer respectively, $f(.)$ denotes the  element-wise activation function (such as ReLU). 

\subsection{Different Loss function}

In E2E optimization, Eq.(\ref{Eq:DCN}) is not the only choice, we can adopt similar pairwise loss (e.g., pairwise logistic loss) shown below:
\begin{eqnarray}
\label{Eq:ps_log_loss}
\mathcal{L^{\text{logis}}}(q,D,Y) = \sum_{q} \sum_{i, j}  {\rm I}(y_i > y_j)  \log \Big[ 1+  \exp ^  {- ( F(q,d_i) - F(q, d_j) ) } \Big] 
\end{eqnarray}


\end{document}
\endinput